\documentclass[twocolumn,showpacs,aps]{revtex4}
\usepackage[dvips]{graphicx}
\usepackage{bm}
\usepackage{amsmath}
\usepackage{amssymb}

\begin{document}
\date{\today}
\title{Theoretical calculation of the surface energy of water}
\author{{Naohisa Ogawa and Yoshiki Matsuura}
\footnote{ogawanao@hit.ac.jp, matsuura@protein.osaka-u.ac.jp}}
\affiliation{Hokkaido Institute of Technology, Maeda, Teine-ku, Sapporo 006-8585 Japan, \\
Institute for Protein Research, Osaka University, Yamada-oka, Suita, Osaka 565-0871 Japan}

\begin{abstract}
The estimation of the surface tension of water is theoretically 
dealt on the basis of the dipole molecular model. It is known 
that the experimentally determined surface tension of freshly 
exposed surface has a higher value than the nominal value of 73 
[mN/m]. We calculated the value corresponding to the fresh 
surface where the reorientation of the molecules has not 
occurred.
\end{abstract}

\pacs{68.37.-d}
\maketitle

\section{Introduction}
 The surface tension of water, $\gamma= 73 [mN/m]$,
 is considered to be a static value after the equilibrium of the rearrangement of the molecules 
at surface is attained. 
The structure of surface orientation \cite{Ori} of water molecules is not yet well-defined, 
though the simulations using molecular dynamics \cite{MD} have been studied. 
This may be one reason that the theoretical calculation of surface tension of water which explains 
the static value has not yet been reported. 
Theoretical treatments of the surface tension have been done for nonpolar van der Waals molecules 
by statistical mechanics, or by using intermolecular forces involving Hamaker constant 
on the basis of the Lifshitz theory\cite{Surfaceenergy},\cite{Jacob}. 

These results showed considerably good agreement with experimental values of nonpolar materials. 
However, theoretical calculation for the polar molecules seems to be hampered by the difficulty 
of evaluation of surface orientation. 
On the other hand, the surface tension of water is known to have higher value than the static one 
for freshly created surface. This has been shown by the dynamical measurement by the oscillating jet method \cite{Jet},
where a value of $180 [mN/m]$ is observed immediately after the surface is created, 
and gradually lowers to its static value within less than $10[mS]$ \cite{kochurova}. 

The higher value is thought to originate from the freshly exposed water surface 
where the molecules are not completely reoriented. We here focused on the calculation of the unoriented 
surface tension of water based on the dipole interaction model. 
The calculation of the contribution of additional terms of intermolecular forces are also presented.

   The surface energy which is defined as the (positive) energy per area of the surface, 
is equivalent to the energy for creation of the surface.
The surface energy originates from the binding energy between molecules.
One molecule inside bulk is surrounded by other molecules.
Let us define the number of such molecules (number of nearest neighbour) as $N_i$, 
and the mean binding energy $W~(<0)$ between nearest two molecules.
Then the binding energy per one molecule is

$$E_i = \frac{N_i}{2}W.$$

 On the other hand, the molecule on the surface is surrounded by less neighbours $N_s$.
Then the binding energy per molecule is

$$E_s=\frac{N_s}{2}W.$$

Therefore the molecule on surface has larger energy than that inside bulk.

$$\Delta E =\frac{N_s-N_i}{2}W >0.$$

By using  $N [m^{-2}]$ as molecule number density on surface, we define the surface energy per area.

$$u =\frac{N_s-N_i}{2}  WN.$$

In the above discussion we considered only the nearest neighbour interaction.
However, we must consider the long range force like Van der Waals and dipole-dipole interactions.
For this purpose, our method for calculating the surface energy is as follows.

\begin{enumerate}
\item Consider the virtual flat surface in bulk which divide bulk into two pieces.
\item Calculate the binding energy of two molecules.
\item Sum up the binding energies for all molecule-pairs  across the virtual surface in bulk.
\item The binding energy obtained above divided by two times surface area gives the surface energy per area.
\end{enumerate}

We will calculate the surface energy of water due to above procedure in the following sections.

The surface tension $\gamma$ which we observe is defined by

\begin{equation}
\gamma = u -Ts,
\end{equation}

where $u$ is the surface energy (per area), and $s$ is the entropy density per area.
This shows our calculation does not directly connect to observable $\gamma$.
However, the surface entropy term is far possible to take into account our consideration, 
since it deeply relates with surface orientation.
We just assume here the surface entropy term is much smaller than the surface energy $u$.
Then the surface tension is approximated by surface energy and 
we compare the experimental value of surface tension of water to the surface energy which we calculate.

   Before starting our program, let us consider the dimensional analysis.
The molecule of water has its own dipole moment.\cite{fletcher} 
\begin{equation}
 \mu = 6.471 \times 10^{-30} [Cm].
\end{equation}
The molecular density is
\begin{equation}
n = 3.35 \times 10^{28} [m^{-3}].
\end{equation}
By using these two quantities: $\mu$ and $n$, 
we obtain only one quantity which has dimension of energy density.
\begin{equation}
\frac{\mu^2 n}{4\pi \epsilon_0} \times n^{2/3} \sim 130[mJ/m^2],
\end{equation}
where we have utilised the value $\frac{1}{4\pi \epsilon_0} = 9.01 \times 10^{9}$.
The surface tension of water is the same order as this value.
This shows  such a consideration has possibility to express the nature of water.

\section{Dipole-Dipole Interaction}
The Interaction energy between two electric dipoles has the form.

\begin{equation}
V=\frac{1}{4\pi\epsilon_0} [\frac{\vec{\mu}_A \cdot \vec{\mu}_B}{r^3} 
- 3 \frac{(\vec{\mu}_A \cdot \vec{r})(\vec{\mu}_B \cdot \vec{r})}{r^5}].
\end{equation}

\begin{center}
\includegraphics[width=1.8cm]{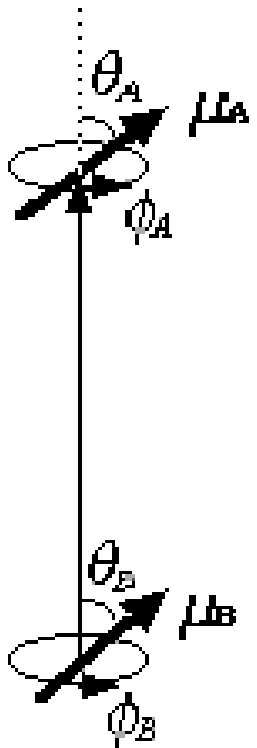}\\
\addtocounter{figure}{1}
Fig.~1
\end{center}

By using the coordinates $\theta_A$, $\theta_B$, $\phi_A$, $\phi_B$ represented in Fig.1, 
This energy is written in the form.

\begin{eqnarray}
V &=& \frac{\mu^2}{4 \pi \epsilon_0 r^3} [\sin\theta_A \sin \theta_B \cos(\phi_A-\phi_B)\nonumber\\
&& ~~~~~  - 2\cos \theta_A \cos \theta_B].
\end{eqnarray}

To obtain the partition function, we further determine the integration measure.  
As two angles $(\theta, \phi)$ specify a point on$S_2$, 
integration measure can be taken as the surface area covering $S_2 \times S_2$, such that
\begin{eqnarray}
Z &=& \int^\pi_0 d\theta_A  \int^\pi_0 d\theta_B \sin \theta_A \sin \theta_B  \nonumber\\
&& \times \int^{2\pi}_0 d\phi_A  \int^{2\pi}_0 d\phi_B \exp (-\beta V).
\end{eqnarray}

Next we will explain how we can obtain surface energy from this function. 
Our purpose is  to obtain the interaction energy between molecules that are placed across the virtual surface.
Let us denote $U$ and $U'$ as  molecules above that surface, and denote $D$ and $D'$ 
as molecules under the surface. This is shown in Fig.2.
Then we need to obtain the thermal expectation value of binding energy between two bulks divided 
by the virtual surface.
\begin{eqnarray}
&<V>& = \prod_{U,D} \int d\mu_U  \int d\mu_D (\sum_{U,D} V_{UD}) \nonumber\\
&&e^{-\beta(\sum_{U \neq U'} V_{UU'} + \sum_{U,D} V_{UD} + \sum_{D \neq D'} V_{DD'})} \nonumber\\
&/& \prod_{U,D} \int d\mu_U \int d\mu_D \nonumber\\
&&e^{-\beta(\sum_{U \neq U'} V_{UU'} + \sum_{U,D} V_{UD} + \sum_{D \neq D'} V_{DD'})},
\end{eqnarray}

where, $d\mu_U$ is the integration measure on the orientation of dipole moment of molecule $U$. 
$V_{UU'}$ is the interaction energy between dipole moments of molecules $U$ and $U'$.
Other notations can be understood in the same way.
We calculate the expectation value of the interaction energy between two dipoles, 
one is in $U$ and another one is in $D$. 
Then we use the mean field approximation by rewriting unremarked other dipole moments to their mean values.
Since there is no specific direction and we know that the water has no spontaneous electric polarisation,
we take these values to be zero.

\begin{center}
\includegraphics[width=4.5cm]{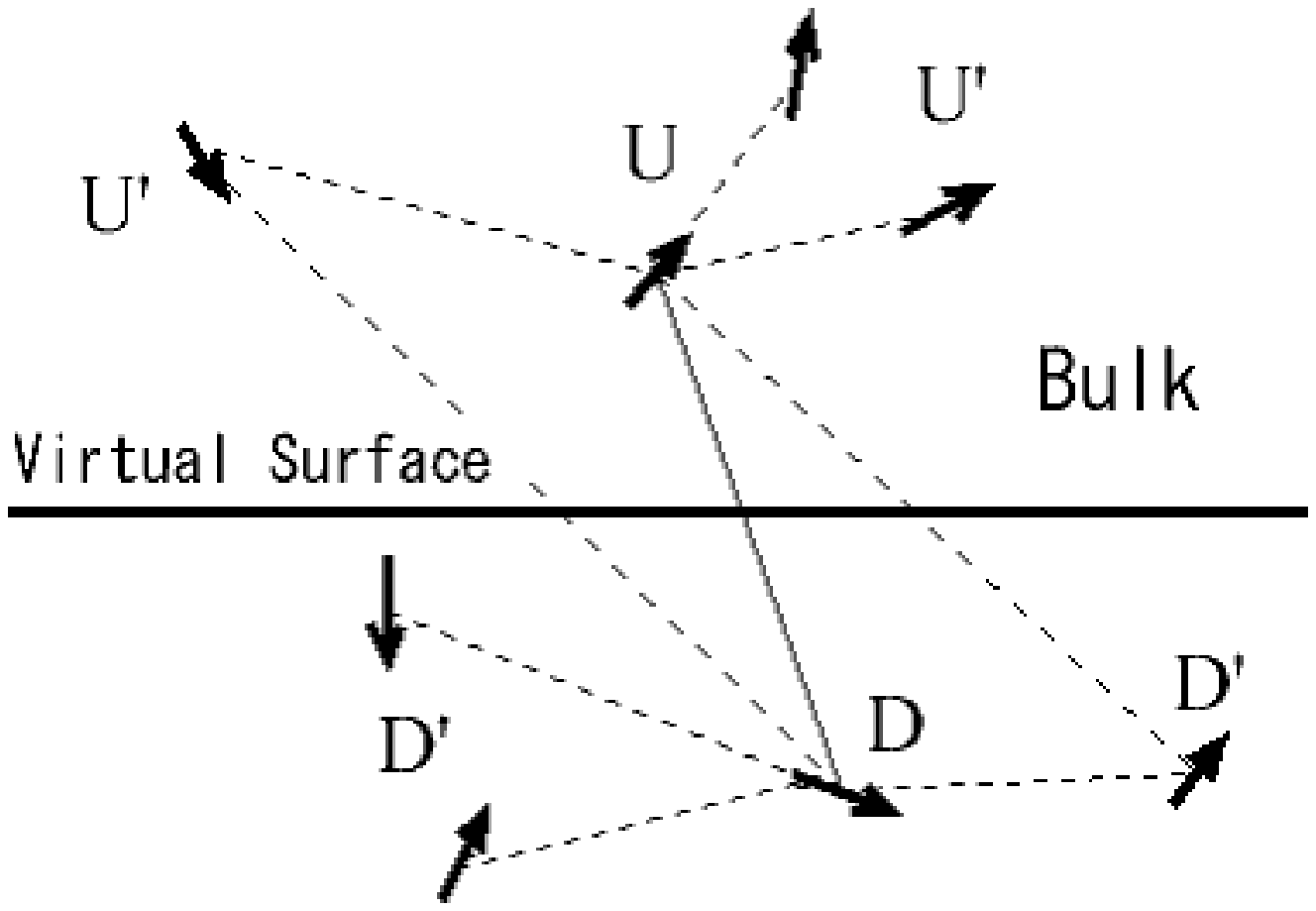}\\
\addtocounter{figure}{2}
Fig.2~~ $<V_{UD}>$: dotted interactions are neglected.
\end{center}
\begin{center}
\includegraphics[width=4.5cm]{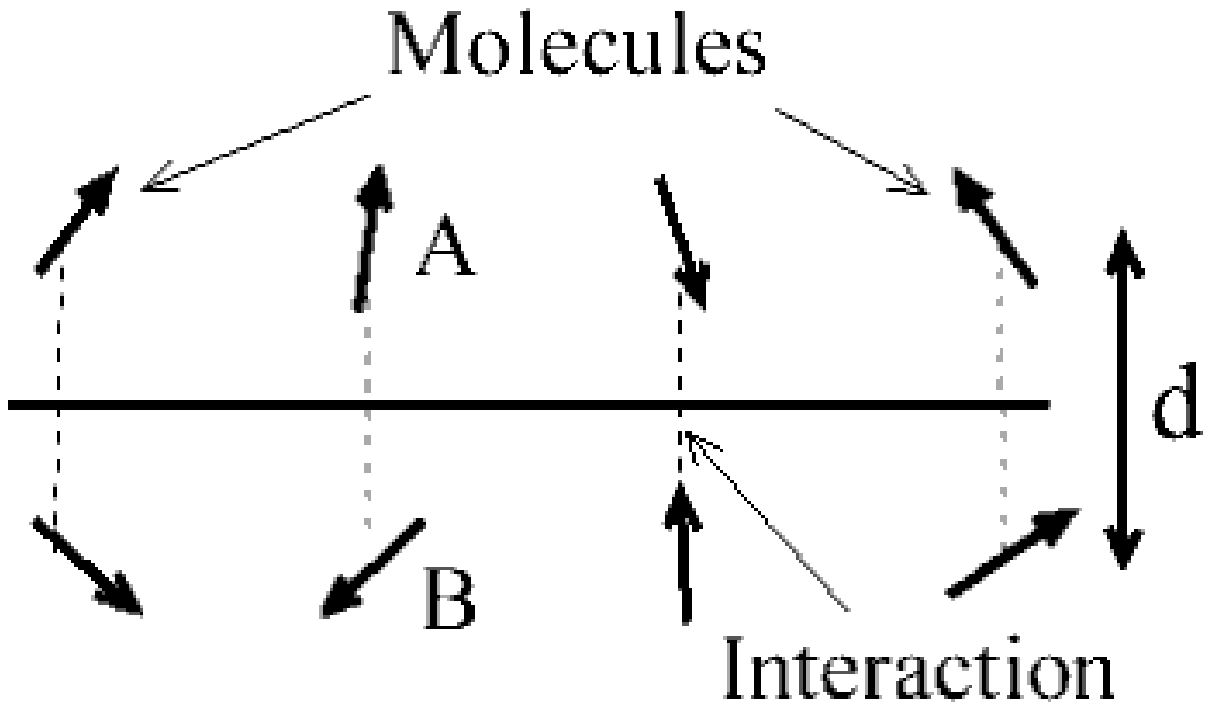}\\
\addtocounter{figure}{3}
Fig.3~~nearest neighbour interactions. $d \equiv n^{-1/3}$: molecular distance.
\end{center}

Then we have
\begin{eqnarray}
<V>&\sim& \sum_{pair:U,D} \int d\mu_U \int d\mu_D  V_{UD}  e^{-\beta V_{UD}} \nonumber\\
&& ~~~~~~~/ \int d\mu_U \int d\mu_D e^{-\beta V_{UD}} \nonumber\\
&& \equiv \sum_{pair:U,D} <V_{UD}>. \label{eq:int}
\end{eqnarray}

In a meanwhile we utilise Eq. (\ref{eq:int})  only for surface molecules like in Fig.3, 
and we calculate the surface energy in this nearest neighbour approximation.
Then the surface energy $U_{sur}$ is given by $<V_{AB}>$ as,
\begin{equation}
U_{sur} = -\frac{<V_{AB}>}{2}N,
\end{equation}
where the negative sign is to make the surface energy positive.
$N$ is the surface density of molecules expressed by $n$.
$$N = \frac{1}{d^2} = n^{2/3}.$$

Further $<V_{AB}>$ is expressed by the partition function.
\begin{equation}
Z= \int d\mu_A \int d\mu_B  \exp[-\beta V_{AB} ],
\end{equation}
\begin{equation}
<V_{AB}> = -\frac{\partial log Z}{\partial \beta}.
\end{equation}
Then we obtain
\begin{equation}
U_{sur} = \frac{1}{2} n^{2/3} \frac{\partial log Z}{\partial \beta}.\label{eq:tension}
\end{equation}

Our approximations are the followings.

\begin{enumerate}
\item Rough mean field approximation.
\item Only the nearest neighbour interaction.
\item Neglecting entropy term for the surface tension.
\end{enumerate}

\section{Calculation of Partition function}
By calculating
\begin{eqnarray}
&&Z= \int^\pi_0 d\theta_A  \int^\pi_0 d\theta_B \sin \theta_A \sin \theta_B  \nonumber\\
&& \times \int^{2\pi}_0 d\phi_A  \int^{2\pi}_0 d\phi_B ~  \exp (-\beta V),
\end{eqnarray}
and by putting it into Eq. (\ref{eq:tension}), we obtain the surface energy in this rough approximation.
First we carry out the $\phi$ integration.
Let us define
\begin{equation}
I_0(\rho) \equiv \frac{1}{(2\pi)^2}\int^{2\pi}_0 d\phi_A  \int^{2\pi}_0 d\phi_B  \exp[-\rho\cos(\phi_A-\phi_B)],
\end{equation}
with
\begin{equation}
\rho \equiv \frac{\beta \mu^2}{4 \pi \epsilon_0 r^3} \sin \theta_A \sin \theta_B.
\end{equation}

We have
\begin{equation}
I_0(\rho) = \frac{1}{\pi} \int_0^{\pi} e^{-\rho \cos \phi} d\phi =  J_0(i\rho),
\end{equation}

where $J_0$ is the 0-th Bessel function, and so $I_0(\rho)$ is the modified Bessel function.
$I_0(\rho)$ can be expanded as,
\begin{equation}
I_0(\rho) = \sum_{m=0}^\infty  \frac{(\rho/2)^{2m}}{(m!)^2}.
\end{equation}
From $\mid \rho \mid \le \frac{\mu^2 n \beta}{4\pi\epsilon_0}\sim 3.4$, 
this expansion converges in the normal temperature.
And so the higher terms can be neglected approximately. By using this relation, we have

\begin{eqnarray}
Z&=& (2\pi)^2 \int^\pi_0 d\theta_A  \int^\pi_0 d\theta_B \sin \theta_A \sin \theta_B  \nonumber\\
&& \hspace{-0.5cm} \times  I_0(\frac{\beta \mu^2}{4 \pi \epsilon_0 r^3} \sin \theta_A \sin \theta_B) e^{2\beta \frac{\mu^2}{4 \pi \epsilon_0 r^3} \cos \theta_A \cos \theta_B}.
\end{eqnarray}
Then  we define
\begin{equation}
\alpha \equiv 2\beta \frac{\mu^2}{4 \pi \epsilon_0 r^3}.
\end{equation}
For $r=d=n^{-1/3}$, $\alpha(T=273K)=6.70,~~\alpha(T=300K)=6.09,~~\alpha(T=373K)=4.90$.
By using the expansion of modified Bessel function, we obtain
\begin{eqnarray}
Z &=& (2\pi)^2  \sum_{m=0}^\infty \frac{(\alpha/4)^{2m}}{(m!)^2}  \int^\pi_0 d\theta_A  \int^\pi_0 d\theta_B \nonumber\\
 &&\times  \sin^{2m+1} \theta_A  \sin^{2m+1} \theta_B  e^{\alpha \cos \theta_A \cos \theta_B}.
\end{eqnarray}

By setting $x=\cos \theta_A, y=\cos \theta_B$, we obtain

\begin{eqnarray}
Z &=& (2\pi)^2  \sum_{m=0}^\infty \frac{(\alpha/4)^{2m}}{(m!)^2}  \int^{+1}_{-1} dx  \int^{+1}_{-1} dy  \nonumber\\
&& \times  (1-x^2)^{m}  (1-y^2)^{m}  e^{\alpha xy}.
\end{eqnarray}

We start from y-integration.
\begin{eqnarray}
&&\int^{+1}_{-1} dy  (1-y^2)^{m}  e^{\alpha xy} \nonumber\\
&&= \sum_{k=0}^m \begin{pmatrix} m \\ k \end{pmatrix}  \int^{+1}_{-1} dy  (-y^2)^{k}  e^{\alpha xy} \nonumber\\
&=& 2 \sum_{k=0}^m \begin{pmatrix} m \\ k \end{pmatrix} (-1)^k \alpha^{-2k}  \frac{d^{2k}}{dx^{2k}} \frac{\sinh(\alpha x)}{\alpha x}.
\end{eqnarray}
Then the partition function becomes
\begin{eqnarray}
Z &=& 2 (2\pi)^2  \sum_{m=0}^\infty Z_m(\alpha),\\
Z_m(\alpha)&\equiv& \frac{(\alpha/4)^{2m}}{(m!)^2}  \sum_{k=0}^m \begin{pmatrix} m \\ k \end{pmatrix} (-1)^k \alpha^{-2k} \nonumber\\
&& \int^{+1}_{-1} dx \quad (1-x^2)^{m}  \frac{d^{2k}}{dx^{2k}} \frac{\sinh(\alpha x)}{\alpha x}.
\end{eqnarray}
And we look for the perturbative solution.
\begin{equation}
Z_0(\alpha)= P_0,
\end{equation}
\begin{equation}
Z_1(\alpha)= (\frac{\alpha}{4})^2 [P_1 -\frac{1}{\alpha^2} P_2],
\end{equation}
\begin{equation}
Z_2(\alpha) = \frac{1}{4}(\frac{\alpha}{4})^4 [P_3 - \frac{2}{\alpha^2} P_4 + \frac{1}{\alpha^4} P_5],
\end{equation}
where,
\begin{eqnarray}
P_0 &\equiv& \int^{+1}_{-1} dx   \frac{\sinh(\alpha x)}{\alpha x} = \frac{2}{\alpha}Shi(\alpha),\\ 
P_1 &\equiv& \int^{+1}_{-1} dx   (1-x^2) \frac{\sinh(\alpha x)}{\alpha x}\nonumber\\
&=& \frac{2}{\alpha}Shi(\alpha)+\frac{2}{\alpha^3}sinh(\alpha)- \frac{2}{\alpha^2}cosh(\alpha) \\
P_2 &\equiv& \int^{+1}_{-1} dx  (1-x^2) \frac{d^{2}}{dx^{2}} \frac{\sinh(\alpha x)}{\alpha x}\nonumber\\
&=& -\frac{4}{\alpha}Shi(\alpha)+\frac{4}{\alpha}sinh(\alpha)\\
P_3 &\equiv& \int^{+1}_{-1} dx  (1-x^2)^2 \frac{\sinh(\alpha x)}{\alpha x}\nonumber\\
 &=& \frac{2}{\alpha}Shi(\alpha) -\frac{2}{\alpha^3}(1+\frac{2}{\alpha^2}) sinh(\alpha) \nonumber\\
 &&  - \frac{2}{\alpha^2}(1-\frac{2}{\alpha^2}) cosh(\alpha), 
\end{eqnarray}
\begin{eqnarray}
P_4 &\equiv& \int^{+1}_{-1} dx   (1-x^2)^2 \frac{d^{2}}{dx^{2}} \frac{\sinh(\alpha x)}{\alpha x}\nonumber\\
&=& -\frac{8}{\alpha}Shi(\alpha)-\frac{24}{\alpha^3}sinh(\alpha) +\frac{24}{\alpha^2}cosh(\alpha), \\
P_5 &\equiv& \int^{+1}_{-1} dx   (1-x^2)^2 \frac{d^{4}}{dx^{4}} \frac{\sinh(\alpha x)}{\alpha x}\nonumber\\
&=& \frac{48}{\alpha}Shi(\alpha)-\frac{64}{\alpha}sinh(\alpha) + 16 cosh(\alpha),
\end{eqnarray}
where $Shi(\alpha)$ is defined by
\begin{equation}
Shi(\alpha)=\int_0^{\alpha} \frac{sinh z}{z} dz=\alpha +\frac{\alpha^3}{3\cdot3!}+\frac{\alpha^5}{5\cdot5!}+\cdots.
\end{equation}

Therefore, up to second order we have
\begin{eqnarray}
&&Z_{(2)}/(2\pi)^2 \equiv 2\sum_{n=0}^{2} Z_{n} = (\frac{147}{32\alpha}+ \frac{9\alpha}{32}+\frac{\alpha^3}{256})Shi(\alpha)\nonumber\\
&&- (\frac{74}{256\alpha}+\frac{\alpha}{256})sinh(\alpha) -(\frac{78}{256} +\frac{\alpha^2}{256})cosh(\alpha) .\label{eq:partition}
\end{eqnarray}

Note that if we expand $V(r)=-\partial ln Z/ \partial \beta$ by $\alpha$, we have 
\begin{equation}
<V(r)>_{lowest}  = -kT \frac{21}{128} \alpha^2 \sim -\frac{2}{3}\frac{\mu^4}{kT (4\pi\epsilon)^2 r^6}.
\end{equation}
This expression of the lowest order in $\alpha$ is well known in many text books as weak field approximation. \cite{adamson},\cite{Jacob}
But in short range interaction ($r \sim d$), $\alpha$ expansion does not converge and such expression fails. 
In this sense our approximation(expansion of modified Bessel function) is essentially different from such a weak field approximation.

Now we obtain surface energy $U_{sur}$ and its temperature gradient$dU_{sur}/dT$ up to second order 
in the following.
\begin{equation}
U^{(2)}_{sur} = \frac{1}{2} n^{2/3} \frac{\partial log Z_{(2)}}{\partial \beta}=
\frac{\mu^2 n^{5/3}}{4\pi\epsilon_0}f^{(2)}(\alpha),
\end{equation}
where
\begin{equation}
f^{(2)}(\alpha) \equiv \frac{1}{Z_{(2)}}\frac{dZ_{(2)}}{d\alpha}.
\end{equation}
\begin{equation}
\frac{dU^{(2)}_{sur}}{dT} = -2k (\frac{\beta \mu^2 n}{4\pi\epsilon_0})^2 n^{2/3} \frac{df^{(2)}}{d\alpha}.
\end{equation}

 The surface energy $U^{(2)}_{sur}$ and its temperature gradient $dU^{(2)}_{sur}/dT$ at normal temperature
 $300K (\alpha = 6.14)$ is
\begin{equation}
U^{(2)}_{sur} = 80.1 [mJ/m^2], ~~ dU^{(2)}_{sur}/dT = -0.108 [mJ/m^2K].
\end{equation}
These values are comparable to the static values of the surface tension and its gradient,
$\gamma_{exp} = 73 [mJ/m^2]$, $(d\gamma/dT)_{exp} = -0.15 [mJ/m^2K].$

Though this result looks good agreement with static value of surface tension,
 it is reported that, in experiments using oscillating jet method \cite{Jet} 
, the value of surface tension of water is about $180[mJ/m^2]$ just after the creation of new surface \cite{kochurova}.
And its value relaxes to the static value $73[mJ/m^2]$ within less than $10[mS]$.
This phenomenon is usually explained by the surface orientation. 
The water molecules on the surface rotate and to make its surface free energy (surface tension) minimum 
during this period. Since we can not take into account the surface orientation from technical reason, 
we need to improve the approximation beyond the nearest-neighbour's one 
and to obtain the value $180[mJ/m^2]$ as the surface tension. 
This is discussed in the next section.

\section{Beyond the Nearest-Neighbour Approximation}
As we have seen in the previous section, the form of binding energy is complicated 
even for one molecule-pair interaction, 
and it seems quite difficult to consider the theory beyond the nearest neighbour interactions.
Furthermore, since the potential energy between two dipoles has the form of  $1/r^3$,
 the sum of the potential energies diverges logarithmically.
Therefore we give an important hypothesis, that is, we utilise thermal expectation value of dipole-dipole 
interaction energy as an elementary potential energy, 
and to sum up these energies for all the molecule pairs across the surface (See Fig.5).\\
   From this hypothesis, the potential depends only on the distance of dipole pair. 
The function of potential starts from $1/r^{6}$, and so we have no divergence in summing up.

  From the previous calculation, we see that the partition function of dipole-dipole interaction 
can be expressed by even function of $\alpha$. 
This means the expectation value of interaction energy: $<V(r)> =-\partial ln Z/ \partial \beta$
 has the distance dependence in the series of $1/r^{6m}$$(m=1,2,3,\dots)$, 
since $\alpha = const. \times \beta/r^3$.\\

Therefore we can write
\begin{equation}
<V(r)> = kT \sum_{m=1}^\infty B_m \alpha^{2m} \equiv \sum_{m=1}^\infty C_m r^{-6m}.
\end{equation}
For the explicit form of $B_m$ and $C_m$, see appendix 1.
By using above relation surface energy in nearest-neighbour approximation can be expressed as,
\begin{equation}
U_{sur}^{N.N}=-\frac{1}{2} n^{2/3} <V(d)> =- \frac{1}{2} n^{2/3} \sum_{m=1}^\infty C_m d^{-6m}.\label{eq:NN}
\end{equation}

Next we consider the case that all the molecules in a bulk interacts with all the ones in another bulk.

\begin{center}
\includegraphics[width=5cm]{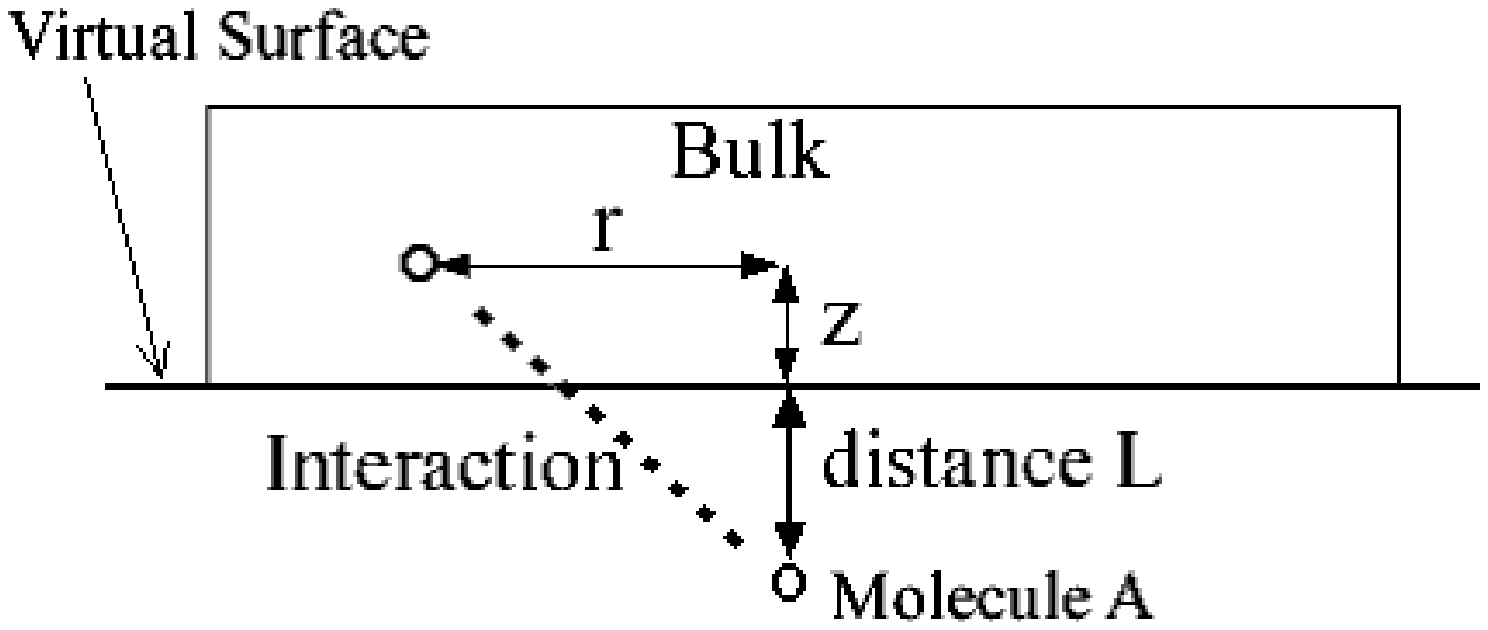}\\
\addtocounter{figure}{4}
Fig.4
\end{center}

\begin{center}
\includegraphics[width=5cm]{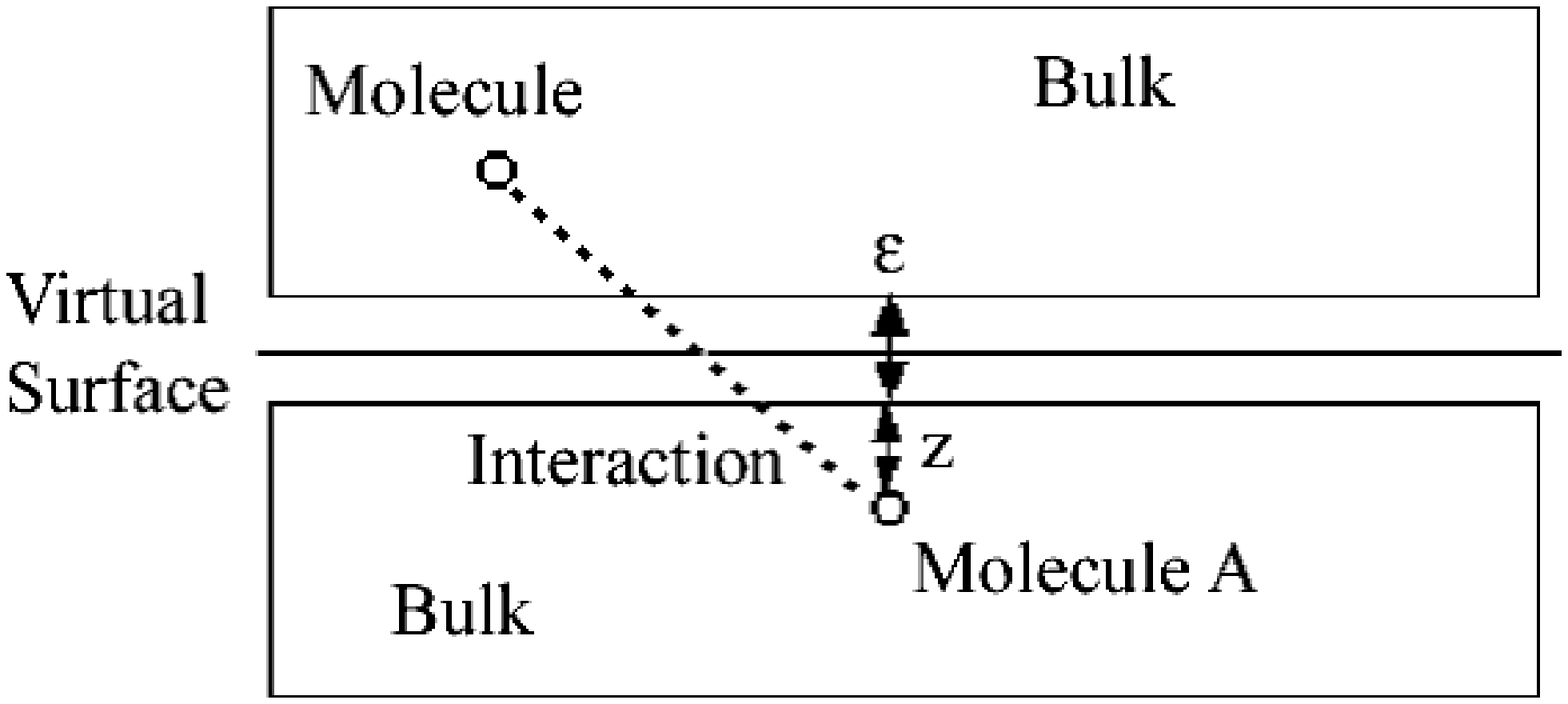}\\
\addtocounter{figure}{5}
Fig.5
\end{center}

First as in Fig.4, the total energy that the molecule $A$ apart $L$ from the surface interacts with
all the molecules in bulk is,
\begin{eqnarray}
U(L) &=& n\int_0^\infty dr 2\pi r \int_0^\infty dz <V(\sqrt{r^2+(z+L)^2})> \nonumber\\
&=& n \pi \sum_{m=1}^\infty \frac{C_m}{(3m-1)(6m-3)} L^{-6m+3}.
\end{eqnarray}
Second as in Fig.5,  we sum up the energy $U(L)$ by changing the position of molecule $A$ in opposite bulk.
This summation is also calculated by using continuous approximation, i.e. by integration.
The distance of two bulks $\epsilon$ is an order of $d$, and its value is quite critical to the result.
So we discuss its value later. Then we have the ``bulk to bulk" surface energy.

\begin{eqnarray}
&&\hspace{-0.6cm}U_{sur}^{B.B} =-\frac{1}{2} n \int_0^\infty dz U(\epsilon+z) \nonumber\\
&&\hspace{-0.6cm}= -\frac{\pi n^2}{2} \sum_{m=1}^\infty \frac{C_m}{(3m-1)(6m-3)(6m-4)} \epsilon^{-6m+4}.\label{eq:BB}
\end{eqnarray}

For $\epsilon=d$, this surface energy is smaller than the one in the case of nearest neighbour approximation, 
though it should be larger. (compare Eq. (\ref{eq:NN}) with Eq. (\ref{eq:BB}) by noting $n=d^{-3}$.)
The reason is the following. Since we have approximated all the molecules are continuously spread out,
the bulk to bulk distance has increased effectively. This is shown in Fig.6.  
Each molecule is spread out continuously in $V_1$ and $V_2$ as shown in (b).
For $\epsilon=d$, the mean distance between two molecules is larger than $d$.
  To determine the virtual bulk-bulk distance $\epsilon$, our consideration follows.

\begin{center}
\includegraphics[width=5cm]{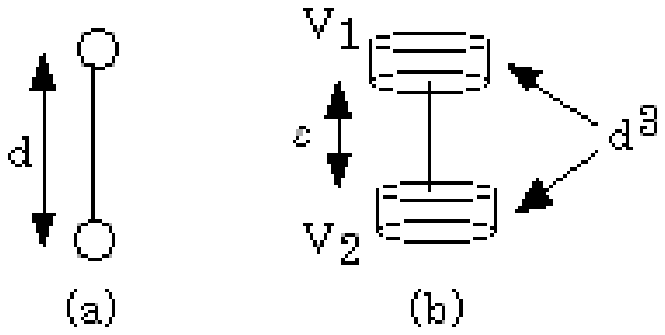}\\
\addtocounter{figure}{6}
Fig.6
\end{center}

 We require that the potential $1/d^{6m}$ as in Fig.6 (a) coincides with the mean potential
 $<1/r^{6m}>$ as in Fig.6 (b), where each molecule is spread out uniformly in column $V_1$ and $V_2$ 
with hight $d$ and basal area $d^2$.
Then we can determine $\epsilon$ for each number of $m$.
\begin{eqnarray}
&&\hspace{-1.0cm}V_d^{(m)} \equiv \frac{1}{d^{6m}} \nonumber\\
&&\hspace{-1.0cm} \sim ~ 
V_\epsilon^{(m)} \equiv \frac{1}{V_1 V_2} \int_{V_1}d^3r_1 \int_{V_2}d^3r_2 \frac{1}{\mid \vec{r}_1-\vec{r}_2 \mid^{6m}}.
\end{eqnarray}
Though it might seem strange that $\epsilon$ depends on $m$, 
the bulk distance $\epsilon$ is virtual mathematical object,
 and it improves the errors coming from continuous approximation at short distance. 
Therefore it can be varied by the parameter of potential form $m$.
Hereafter we rewrite $\epsilon \to \epsilon_m$.
As we have shown in appendix 2, in large $m$ limit this condition reduces to
\begin{equation}
\frac{1}{d^{6m-4}} \sim \frac{\pi}{(3m-1)(6m-3)(6m-4)}\frac{1}{\epsilon_m^{6m-4}}.\label{eq:LLm}
\end{equation}
If we compare Eq. (\ref{eq:NN}) with Eq. (\ref{eq:BB}) by using above relation, 
we see the coincidence of the series expansion of $U^{BB}$ and $U^{NN}$ at large $m$.
This is  not surprising. For large $m$,  the short range interaction enhanced, 
and the nearest neighbour interaction plays a central role in surface energy.
For $m=1,2$ we obtain in appendix 2,
$$\epsilon_1=0.373 d,~~~\epsilon_2=0.54639 d.$$
And for $m \ge 3$,  Eq. (\ref{eq:LLm}) approximately holds and we just rewrite the series expansion of $U^{BB}$ by using  $U^{NN}$.

\begin{eqnarray}
U_{sur}^{N.N}&=&-\frac{1}{2} n^{2/3} \sum_{m=1}^\infty C_m d^{-6m},\nonumber\\
U_{sur}^{B.B} &=&-\frac{1}{2} n^{2} \pi  [ \frac{C_1}{12} \epsilon_1^{-2} +\frac{C_2}{360} \epsilon_2^{-8}] \nonumber\\
&&- \frac{1}{2} n^{2/3} \sum_{m=3}^\infty C_m d^{-6m}
\end{eqnarray}
We should remark here that these expansions are not convergent for real value of d.
But we had to expand the potential energy in powers of $r$ to sum up  the interaction energies for all the molecule-pairs across the virtual surface.
Though our two expansions are not convergent, the difference of these two surface energies is convergent.
By using the value of $\epsilon$ which are calculated in appendix 2, we have
\begin{eqnarray}
U_{sur}^{B.B} -U_{sur}^{N.N}= -\frac{1}{2} n^{2/3}  [ \frac{0.8817 C_1}{d^6} +\frac{0.09856 C_2}{d^{12}}] 
\end{eqnarray}
The r.h.s. is the correction to nearest-neighbour approximation, which comes from distant interactions. 
From $C_m/d^{6m} = kT B_m \alpha^{2m}$, we have 
\begin{eqnarray}
U_{sur}^{B.B} &-&U_{sur}^{N.N} = -\frac{n^{2/3}}{2\beta} [ 0.8817 B_1\alpha^2 +0.09856 B_2 \alpha^4] \nonumber\\
&=& -\frac{\mu^2 n^{5/3}}{4\pi \epsilon_0} [ 0.8817 B_1\alpha +0.09856 B_2 \alpha^3]\nonumber\\
&=& -131[ 0.8817\times (-\frac{21}{128})\alpha \nonumber\\
&& +0.09856 \times (\frac{9779}{2457600}) \alpha^3] [mJ].
\end{eqnarray}

Since we know the analytical value of $U^{NN}_{sur}$, all the surface energy due to dipole-dipole interaction becomes
\begin{eqnarray}
&&U^{B.B}_{sur} = \frac{\mu^2 n^{5/3}}{4\pi \epsilon_0} [f(\alpha)-( 0.8817\times (-\frac{21}{128})\alpha  \nonumber\\
&&+0.09856 \times (\frac{9779}{2457600}) \alpha^3)]\nonumber\\
&& = 131 \times [f(\alpha)+0.14465\alpha-0.0003922\alpha^3],
\end{eqnarray}
where the unit is given by $[mJ]$.
The Fig.7 shows the $\alpha$ dependence of the surface energy. 
\begin{center}
\includegraphics[width=4.5cm]{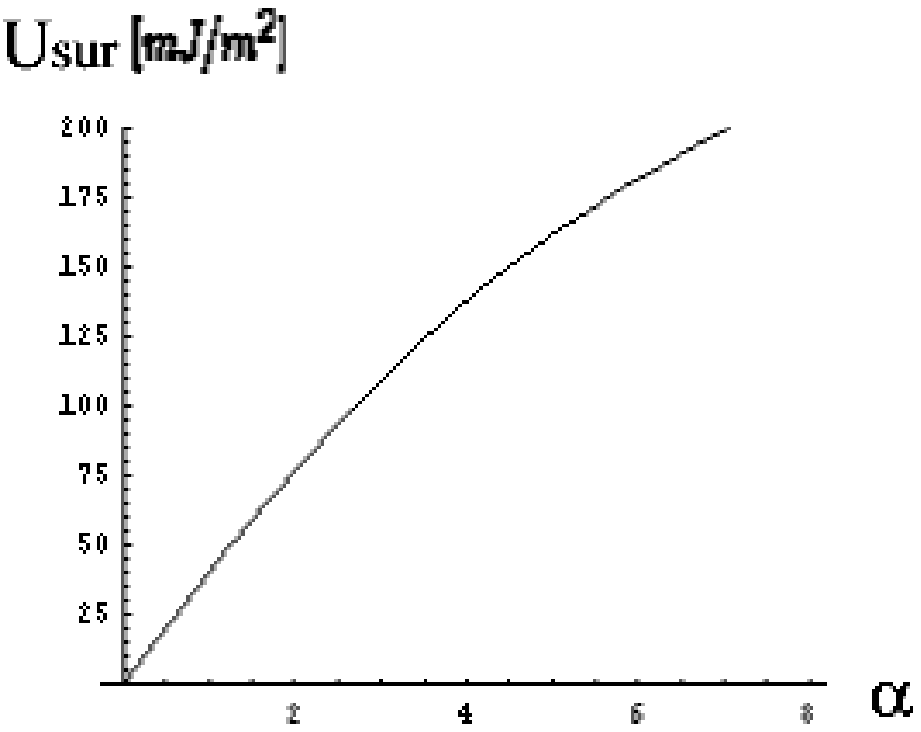}\\
\addtocounter{figure}{7}
Fig.7~~$U_{sur}-\alpha$
\end{center}

Our theory gives the surface energy $185[mJ/m^2]$ at normal temperature $(300K)$.
This result well coincides with the experimental value $180[mJ/m^2]$ 
by using the oscillating jet method \cite{Jet}, \cite{kochurova}.
However, we need to take into account other interactions. 
This is done in next section.

\section{Additional Other Forces}
We still have 3 kind of forces that work between molecules such as, dispersion force,  induced force, 
and  repulsive force  (part of Lennard-Jones potential). 
So we must include them into our consideration of surface energy.

 The dispersion force is essentially quantum mechanical effect.\cite{Jacob}
\begin{equation}
U_{disp} = -\frac{3}{4}(\frac{\alpha_0}{4\pi\epsilon_0})^2 \frac{I}{r^6},
\end{equation}
where $\alpha_0$, the electronic polarisation rate and  $I$, the first ionization energy.
These values are given as $\alpha_0/(4\pi\epsilon_0)=1.48\times 10^{-30}[m^3],~~I=12.6[eV]$.\\

The induced force is the interaction between static molecular dipole and 
the instantaneous dipole moment due to the fluctuation of electron cloud.\cite{Jacob}

\begin{equation}
U_{ind} = -\frac{2 \mu^2 \alpha_0}{(4\pi\epsilon_0)^2 r^6}.
\end{equation}

The experimental value of the repulsive energy,  a part of Lennard-Jones potential is, \cite{Reid},\cite{Lide}
\begin{equation}
U_{rep}=4\varepsilon (\frac{\sigma}{r})^{12},
\end{equation}
where $\varepsilon/k=809.1[K],~~\sigma=2.641\times 10^{-10}[m]$.\\

 We sum up  these 3 energies by using $d=n^{-1/3}=3.224\times 10^{-10}[m]$,

\begin{eqnarray}
&&\Delta U \equiv U_{disp} + U_{ind} + U_{rep} \nonumber\\
&&= [-4.97 (\frac{d}{r})^6 + 4.08 (\frac{d}{r})^{12}] \times 10^{-18} [mJ].
\end{eqnarray}
Now our consideration goes to the surface energy as before.
We sum up all the interaction energies across the virtual interface.
For the molecule-molecule potential
$$V(r) = \sum_{m=1}^{\infty} \frac{C_m}{r^{6m}},$$
Bulk-bulk surface energy becomes
$$U_{sur} = -\frac{\pi}{2}n^2 \sum_{m=1}^{\infty} \frac{C_m}{(3m-1)(6m-3)(6m-4)} \epsilon_m^{-6m+4},$$
where $\epsilon_1=0.373d,~~\epsilon_2=0.54639d$.
 In [mJ] unit we obtain for this additional energy
$$C_1=-4.97 \times 10^{-18} d^6, ~~~ C_2 =4.08 \times 10^{-18} d^{12}.$$
Then we have additional surface energy due to these 3 forces.
\begin{eqnarray}
&&\Delta U_{sur} = -\frac{\pi}{2}n^{2/3} [-\frac{4.97}{12} (\frac{1}{0.373})^2 \nonumber\\
&& + \frac{4.08}{360} (\frac{1}{0.54639})^{8}] \times 10^{-18} = 25[mJ/m^2].
\end{eqnarray}

From above calculations, the total surface energy at normal temperature is
\begin{equation}
U_{sur} = U^{dipole} + U^{disp}+ U^{ind}+ U^{rep} =210.
\end{equation}
Comparing to the experimental value of surface tension: $180[mJ/m^2]$, the obtained value is only 17 \%  
larger. \cite{kochurova}
So we may conclude that our method to obtain the surface tension expresses the nature of water. 

\section{Summary}
We have calculated the dipole-dipole interaction energy due to new approximation by using statistical mechanics.
Then we sum up the energies for all the molecule pairs across the surface in bulk.
In this way we have calculated the surface energy of water as dipoler liquid.
Our treatment is the first step for obtaining the surface energy, though some improvements may be necessary.
The first improvement should be done for the rough mean field approximation.
The thermal expectation value $<V_{UD}>$ may be affected by other environmental molecules.
This might be done by introducing the effect of dielectric constant like the Lifschitz theory.

\section{Appendix 1}
We expand the partition function Eq. (\ref{eq:partition}) by $\alpha$ series.
\begin{eqnarray}
&& Z_{(2)}/(2\pi)^2 =4+\frac{21}{64}\alpha^2 +\frac{91}{9600}\alpha^4 \nonumber\\
&&+\frac{113}{806400}\alpha^6 +\frac{407}{304819200}\alpha^8 +\dots.
\end{eqnarray}

In the same way,
\begin{eqnarray}
&&\frac{<V(r)>}{kT} \sim -\beta \frac{\partial \log Z_{(2)}}{\partial \beta} \nonumber\\
&&= -\frac{21}{128}\alpha^2+\frac{9779}{2457600}\alpha^4  \nonumber\\
&&- \frac{650953}{4404019200}\alpha^6 +\frac{92250213859}{15981304872960000}\alpha^8 \nonumber\\
&&- \frac{7557351481321}{33002459982987264000}\alpha^{10}+\cdots.
\end{eqnarray}
Each coefficient of $\alpha^{2m}$ is $B_m$.
Then the form of $C_m$ is given by
\begin{equation}
C_m = \beta^{-1} B_m (\frac{2\beta \mu^2}{4\pi \epsilon_0})^{2m}.
\end{equation}

\section{Appendix 2}
\begin{center}
\includegraphics[width=4cm]{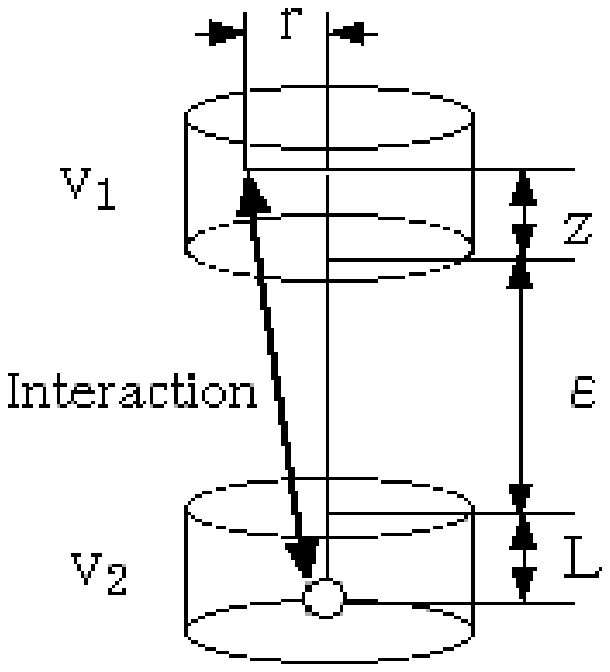}\\
\addtocounter{figure}{8}
Fig.8
\end{center}
\begin{eqnarray}
V_\epsilon^{(m)} &\equiv& \frac{1}{V_1 V_2} \int_{V_1}d^3r_1 \int_{V_2}d^3r_2 \frac{1}{\mid \vec{r}_1-\vec{r}_2 \mid^{6m}}\nonumber \\
&=& \frac{1}{d} \int^d_0 dL~~ \frac{1}{d^3} \int^d_0 dz \nonumber\\
&& \times \int^{d/\sqrt{\pi}}_0 \frac{2\pi r dr}{\sqrt{(\epsilon +L+z)^2+r^2}^{6m}}.
\end{eqnarray}
Since we have a translational invariance to the inter facial direction, when we integrate by 
$\vec{r}_1$ in column $V_1$, we fix $\vec{r}_2$ in the centre axis in column $V_2$.
After that, the integration of $\vec{r}_2$ is given only in the centre axis direction such as, 
integrating from $0$ to $d$ and divided by $d$.\\

  First  r-integration can be done and we obtain
\begin{eqnarray}
V_\epsilon^{(m)} &=& \frac{\pi}{(3m-1)d^4} \int_{0}^d dL \int_{0}^d dz (\frac{1}{(\epsilon+L+z)^{6m-2}}\nonumber\\
&&- \frac{1}{[d^2/\pi +(\epsilon+L+z)^2]^{3m-1}}).
\end{eqnarray}

Next by using the formula
\begin{eqnarray}
&&\int^1_0 dx ~\int^1_0 dy~ g(x+y) \nonumber\\
&&\hspace{-0.6cm} =2\int^2_1 du ~g(u) + \int^1_0 du~ u~ g(u) - \int^2_1 du~ u~ g(u),
\end{eqnarray}
and by setting $x=z/d, ~ y=L/d,~u=x+y,~ \epsilon' \equiv \epsilon/d$,
\begin{eqnarray}
&&V_\epsilon^{(m)} = \frac{\pi}{(3m-1)d^{6m}}\nonumber\\
&&\hspace{-0.4cm} \times [2 \int^{2+\epsilon'}_{1+\epsilon'} du (\frac{1}{u^{6m-2}}-\frac{1}{(1/\pi +u^2)^{3m-1}}) \nonumber\\
&&\hspace{-0.4cm} - \int^{2+\epsilon'}_{1+\epsilon'} du (u-\epsilon') (\frac{1}{u^{6m-2}}-\frac{1}{(1/\pi +u^2)^{3m-1}})\nonumber \\
&&\hspace{-0.4cm} + \int^{1+\epsilon'}_{\epsilon'}du (u-\epsilon') (\frac{1}{u^{6m-2}}-\frac{1}{(1/\pi +u^2)^{3m-1}})].
\end{eqnarray}
In the first and second integration term, the integrand becomes smaller since $u>1$ for large $m$, 
Therefore we have
\begin{eqnarray}
&&V_\epsilon^{(m)} \mid_{m \to \infty} \sim \frac{\pi}{(3m-1)d^{6m}}
 \int^{1+\epsilon'}_{\epsilon'}du (u-\epsilon') \nonumber\\
&&~~~~~~ \times (\frac{1}{u^{6m-2}}-\frac{1}{(1/\pi +u^2)^{3m-1}})\nonumber\\
&&\sim \frac{\pi}{(3m-1)d^{6m}}[\frac{1}{(6m-4)(6m-3)\epsilon'^{6m-4}}\nonumber\\
&&~~~ - \int^{1+\epsilon'}_{\epsilon'}du\frac{u-\epsilon'}{(1/\pi +u^2)^{3m-1}}].
\end{eqnarray}
From the numerical calculation, the rate of the last integration term to the first term is about 8\% for $m=2$ and 2.5\% for $m=3$.
So we can neglect the last integration term  for $m \ge 3$.
Therefore the condition $1/d^{6m}=V_\epsilon^{(m)}$ is simplified in large $m$ limit as
\begin{equation}
 \frac{(3m-1)(6m-3)(6m-4)}{\pi} \sim \frac{1}{\epsilon'^{6m-4}} \label{eq:Lm}
\end{equation}
If we compare Eq. (\ref{eq:NN}) with Eq. (\ref{eq:BB}) using above relation, 
we see the coincidence of the series expansion of $U^{BB}$ and $U^{NN}$ at large $m$.
This is  quite natural result. 
Because at large $m$, only the short range interaction enhanced,
 and so the nearest neighbour interaction gives the main effect.

The condition $1/d^{6m}=V_\epsilon^{(m)}$ is now rewritten as
\begin{eqnarray}
&&\frac{(3m-1)(6m-3)(6m-4)}{\pi}-\frac{1}{\epsilon'^{6m-4}} \nonumber\\
&&= \frac{1}{(2+\epsilon')^{6m-4}}-\frac{2}{(1+\epsilon')^{6m-4}} \nonumber\\
&&- (6m-3)[\frac{1}{((2+\epsilon')^2 +1/\pi)^{3m-2}} \nonumber\\
&&-\frac{2}{((1+\epsilon')^2 +1/\pi)^{3m-2}}+\frac{1}{(\epsilon'^2 +1/\pi)^{3m-2}}]\nonumber\\
&&- (6m-3)(6m-4)[(2+\epsilon') \int^{2+\epsilon'}_{1+\epsilon'} du \frac{1}{(u^2 +1/\pi)^{3m-1}}\nonumber\\
&&- \epsilon'\int^{1+\epsilon'}_{\epsilon'} du \frac{1}{(u^2 +1/\pi)^{3m-1}}].
\end{eqnarray}
The remained integrations in right hand side can be calculated in series expansion for general $m$.
For $m=1$, this condition becomes
\begin{eqnarray}
&&\hspace{-0.4cm}\frac{12}{\pi} -\frac{1}{\epsilon'^{2}} = \frac{1}{(2+\epsilon')^{2}}-\frac{2}{(1+\epsilon')^{2}} \nonumber\\
&&\hspace{-0.4cm}+ 3\sqrt{\pi}^3 [ -(2+\epsilon')\arctan(\sqrt{\pi}(2+\epsilon')) \nonumber\\
&&\hspace{-0.4cm}+ 2(1+\epsilon')\arctan(\sqrt{\pi}(1+\epsilon')) - \epsilon' \arctan(\sqrt{\pi}\epsilon')]
\end{eqnarray}
By solving this, we have $\epsilon'_1=0.373$. 
In the same way we obtain $\epsilon'_2=0.54639$, and $\epsilon'_3=0.63718926$.
If we use Eq. (\ref{eq:Lm}) for $m=3$ instead, we have $\epsilon'_3=0.638$.
From above calculation we use above numerical value for $m=1,2$, 
and we use $U^{NN}$ instead of $U^{BB}$ for $m \ge 3$.

\end{document}